\documentclass[pre,floatfix,twocolumn,showpacs,a4]{revtex4}

\usepackage{graphicx}%
\usepackage{dcolumn}
\usepackage{amsmath}
\usepackage{amssymb}
\usepackage{tabularx}
\usepackage{epsfig}
\begin{document}


\title[Short Title]{Generalizations of the clustering coefficient to weighted complex networks}
\author{Jari Saram\"{a}ki$^1$}
\email{jsaramak@lce.hut.fi}
\author{Mikko Kivel\"{a}$^1$}
\author{Jukka-Pekka Onnela$^{1,2}$}
\author{Kimmo Kaski$^1$}
\author{J\'{a}nos Kert\'{e}sz$^{1,3}$}
\affiliation{
$^1$Laboratory of Computational Engineering, Helsinki University of Technology, P.O. Box 9203, FIN-02015 HUT, Finland \\
$^2$Department of Physics, Clarendon Laboratory, University of Oxford, Oxford, OX1 3PU, U.K.\\
$^3$Department of Theoretical Physics, Budapest University of Technology and 
Economics, Budapest, Hungary
}

\date{\today}
\begin{abstract}
The recent high level of interest in weighted complex networks gives rise to 
a need to develop new measures and to generalize existing ones to take
the weights of links into account. Here we focus on various generalizations of 
the clustering coefficient, which is one of the central characteristics in
the complex network theory. We present a comparative study of the several 
suggestions introduced in the literature, and point out their advantages and 
limitations. The concepts are illustrated by simple examples as well as by 
empirical data of the world trade and weighted coauthorship networks.

\end{abstract}
\maketitle

The study of networks has become a central topic in the science of 
complex systems \cite{ABReview,NewmanReview,DorogovtsevReview}. 
In the network approach, interacting elements are depicted as vertices 
in the network and their interactions as edges connecting the vertices. 
The inherent strength of this approach lies in its ability to capture some of 
the essential characteristics of interacting systems by disregarding 
the detailed nature of both the constituents and the interactions between them. 
Studies on structural properties of complex networks have revealed features 
common to a large number of natural and man-made systems, such as short 
average path lengths, broad degree distributions, modularity, and high 
level of clustering. 

Recently, it has become increasingly clear that in order to understand better 
the properties of the system, it is necessary to take into account some of its 
hitherto omitted details. In particular, understanding the heterogeneity of 
interaction strenghts and their correlations with network topology is 
fundamental in studies of several types of networked systems, e.g. social
 and traffic networks. 
This heterogeneity can be taken into account by assigning \emph{weights} to the 
network edges to quantify, e.g.~fluxes in traffic-related networks~\cite{cbarrat} 
(air traffic, Internet), strengths of social ties~\cite{WattsScience}, correlations 
between stock returns~\cite{JP_PRE}, and trade volumes between countries. 

Incorporating this additional degree of freedom in the complex networks framework 
calls for entirely novel measures as well as generalizations of the 
existing ones. Some of these measures are readily generalizable, e.g. the vertex degree 
$k_i$, denoting the number of edges connected to vertex $i$. For this the natural 
weighted counterpart is the vertex \emph{strength} 
$s_i=\sum_{j\in \nu_i} w_{ij}$~\cite{cbarrat}, where $\nu_i$ denotes the 
neighbourhood of $i$ and $w_{ij}$ are the weights of edges emanating from 
$i$~\cite{footnote:undirected}. Unfortunately, not all existing network characteristics
can be generalized in such a straightforward manner. Here we will focus on the
several alternative definitions proposed in the recent literature for the \emph{weighted clustering coefficient}.

A large number of networks show a tendency for link 
formation between neighbouring vertices, i.e. the network topology deviates from 
uncorrelated random networks in which triangles are sparse. This tendency is called 
\emph{clustering}~\cite{SW,Szabo}, 
and it reflects the clustering of edges into 
tightly connected neighbourhoods. Its origins can be traced back to sociology, 
where similar concepts have been used~\cite{Wasserman,Scott} -- in a typical 
social network, the friends of a person are very likely to know each other. The 
clustering around a vertex $i$ is quantified by the (unweighted) clustering coefficient 
$C_i$, defined as the number of triangles in which  
vertex $i$ participates normalized by the maximum possible number of such triangles: 
\begin{equation}
C_i=\frac{2t_i}{k_i\left(k_i-1\right)},
\label{unweighted_c}
\end{equation}
where $t_i$ denotes the number of triangles around $i$. Hence $C_i=0$ 
if none of the neighbours of a vertex are connected, and $C_i=1$ if all of the neighbours are connected. 
In network analysis this quantity can then be averaged over the entire network or by vertex degree. 

By extending the above line of reasoning, 
the weighted clustering coefficient should also take into account how much weight 
is present in the neighbourhood of the vertex, compared to some limiting case. 
Evidently, this can be done in several ways, and in what follows we focus on four 
existing definitions. In all these formulas, $w_{ii}=0$ $\forall$ $i$, i.e., 
self-edges are not allowed, and $j,k \in \nu_i$. 

\emph{-Barrat et al.} were the first to propose a weighted version of the 
clustering coefficient \cite{cbarrat}. Their definition reads as follows:
\begin{equation}
\tilde{C}_{i,B}=\frac{1}{s_i\left(k_i-1\right)}\sum_{j,k}\frac{w_{ij}+w_{ik}}{2}a_{ij}a_{jk}a_{ik}, \label{C_BBV}
\end{equation}
where $a_{ij}=1$ if there is an edge between $i$ and $j$, and 0 otherwise. Noting that 
$s_i=k_i \left(s_i/k_i\right) = k_i \left<w_{i}\right>$, this may also be written as 
\begin{displaymath}
\tilde{C}_{i,B}=\frac{1}{k_i\left(k_i-1\right)}
\sum_{j,k}\frac{1}{\left<w_{i}\right>}\frac{w_{ij}+w_{ik}}{2}a_{ij}a_{jk}a_{ik},
\end{displaymath}
where $\left<w_{i}\right>=\sum_{j} w_{ij}/k_i$.
As the rewritten form shows clearly, the contribution of each triangle is weighted by a ratio
of the average weight of the two adjacent edges of the triangle to the average weight $\left<w_i\right>$.

\emph{-Onnela et al.} proposed a version \cite{intensity} of weighted clustering 
coefficient based on the concept of subgraph \emph{intensity}, defined 
as the geometric average of subgraph edge weights, resulting in:
\begin{equation}
\tilde{C}_{i,O}=\frac{1}{k_i\left(k_i-1\right)}\sum_{j,k}\left(\hat{w}_{ij}\hat{w}_{ik}\hat{w}_{jk}\right)^{1/3}.
\label{C_LCE}
\end{equation}
Here the edge weights are normalized by the maximum weight in the network,
$\hat{w}_{ij}=w_{ij}/max(w)$ and the contribution of each triangle depends on 
all of its edge weights. Thus triangles in which each edge weight equals  
$max(w)$ contribute unity to the sum, while a triangle having one 
link with a negligible weight will have a negligible contribution. 
This definition can be rewritten as 
\begin{equation}
\tilde{C}_{i,O}=C_i \bar{I_i}, {\text{ with }} \bar{I_i}=\frac{1}{2t_i}\sum_{j,k}\left(\hat{w}_{ij}\hat{w}_{ik}\hat{w}_{jk}\right)^{1/3} 
\label{avg_int}
\end{equation}
where $C_i$ is the unweighted clustering coefficient and $\bar{I_i}$
denotes the average (normalized) intensity of triangles in which vertex $i$ participates.

-\emph{Zhang et al.}~have defined, in the context of gene co-expression networks \cite{czhang}, 
the weighted clustering coefficient as 
\begin{equation}
\tilde{C}_{i,Z}=\frac{\sum_{j,k}\hat{w}_{ij}\hat{w}_{jk}\hat{w}_{ik}}
{\left(\sum_{k}\hat{w}_{ik}\right)^2
- \sum_{k}\hat{w}_{ik}^2}, \label{czhang_v1}
\end{equation}
where the weights have again been normalized by $max(w)$ as above.
The logic behind this definition is the following: 
the number of triangles around vertex $i$ can be 
written in terms of the adjacency matrix elements as 
$t_i=\frac{1}{2}\sum_{j,k}a_{ij}a_{jk}a_{ik}$, and the numerator 
of Eq.~(\ref{czhang_v1}) is simply a weighted generalization of this formula. 
The denominator has been chosen by considering the upper bound of the numerator, 
ensuring $\tilde{C}_{i,Z} \in [0,1]$. This formula can also be written 
as ~\cite{Kalna06}
\begin{equation}
\tilde{C}_{i,Z}=\frac{\sum_{j,k}\hat{w}_{ij}\hat{w}_{jk}\hat{w}_{ik}}
{\sum_{j \neq k}\hat{w}_{ij}\hat{w}_{ik}}.
\label{czhang_v2}
\end{equation}
A similar definition has also been presented in 
Refs.~\cite{Grindrod02,Ahnert06}, where the edge weights are
interpreted as probabilities such that in an ensemble of networks, 
$i$ and $j$ are connected with probability $\hat{w}_{ij}$.  

-\emph{Holme et al.}~have defined the weighted clustering coefficient 
as \cite{cholmekim}
\begin{equation}
\tilde{C}_{i,H}  =
\frac{\sum_{j,k}w_{ij}w_{jk}w_{ki}}{max(w)\sum_{j,k}w_{ij}w_{ki}} 
 =  \frac{\mathbf{W}^3_{ii}}{\left(\mathbf{W}\mathbf{W}_{max} \mathbf{W}\right)_{ii}}, 
\end{equation}
where $\mathbf{W}$ denotes the weight matrix, and $\mathbf{W}_{max}$ a matrix
where each entry equals $max(w)$. The lines of reasoning look similar to those of 
Ref.~\cite{czhang}; however, $j \neq k$ is not required in the denominator sum. 

\newcommand{\rb}[1]{\raisebox{-1.5ex}[0pt]{#1}}
\newcommand{\rbb}[1]{\raisebox{-1.8ex}[0pt]{#1}}
\begin{table}[t]

\begin{tabularx}{8cm}{|c|X|}\hline
Coeff. & Motivation \\
\hline \hline

\rb{$\tilde{C}_B$} & Reflects how much of vertex strength is associated with adjacent triangle edges\\
\hline

\rb{{$\tilde{C}_O$}} & Reflects how large triangle weights are compared to network maximum\\
\hline
\rb{$\tilde{C}_Z$} & Purely weight-based; insensitive to additive 
noise which may result in appearance of ``false positive'' edges with small weights\\

\hline
{$\tilde{C}_H$} & Similar to $\tilde{C}_Z$, based only on edge weights
\\
\hline
\end{tabularx}
\\
\vspace{1mm}
\begin{tabularx}{8cm}{|X||c|c|c|c|}\hline
Feature & $\tilde{C}_B$ & $\tilde{C}_O$ & $\tilde{C}_Z$ & $\tilde{C}_H$ \\
\hline\hline
1) $\tilde{C}$ = C when weights become binary & {X} & {X} & X  &  \\
\hline
2) $\tilde{C} \in [0,1]$ & {X} & {X} & X  &  \\
\hline
3) Uses global $max(w)$ in normalization &   & \rb{X} & \rb{X} & \rb{X} \\
\hline
4) Takes into account weights of all edges in triangles &   & \rbb{X} &  & \rbb{X}  \\
\hline
5) Invariant to weight permutation for one triangle &   & \rbb{X} &   &   \\
\hline
6) Takes into account weights of edges not participating in any triangle & \rbb{X} &   & \rbb{X} & \rbb{X} \\
\hline
\end{tabularx}
\caption{Motivation and comparison of selected features for different weighted clustering coefficients.} \label{comp_table}
\end{table}

Table \ref{comp_table} presents the selected features of the four weighted clustering coefficients
and  illustrates their differences. These features are discussed in detail below. 
In Table \ref{comp_table} and in what follows, $\tilde{C}$ denotes the weighted 
clustering coefficient and $C$ the corresponding unweighted coefficient, with properties
summarized below:

\begin{enumerate}
\item $\tilde{C}=C$ when weights are binary, i.e., $w_{ij}=1$ if $i$ and $j$ are connected. 
This condition is fulfilled 
by all weighted clustering coefficients except that of Holme \emph{et al.} When the weights
are set to binary, $\tilde{C}_{i,H}=2t_i/k_i^2$, which approaches the unweighted coefficient
only when $k\gg 1$. 

\item $\tilde{C} \in [0,1]$. This is true for all weighted 
coefficients except $\tilde{C}_{i,H}$, which never reaches unity for the reason mentioned above. 
Let us consider the limiting values ($\tilde{C_i}=0$, $\tilde{C_i}=1$) in more detail. 
For all coefficients, $\tilde{C_i}=0$ signifies the absence of triangles. A necessary 
condition for $\tilde{C}_{i,B}=1$, $\tilde{C}_{i,O}=1$, and $\tilde{C}_{i,Z}=1$ is 
that edges exist between all neighbours of vertex $i$. However, each coefficient 
sets a different requirement for the weights. When $C=1$, then $\tilde{C}_{i,B}=1$ 
irrespective of the edge weights. Contrary to this, $\tilde{C}_{i,O}=1$ requires 
that the weights of all edges $w_{ij}=w_{jk}=max(w)$, i.e., all weights in each 
triangle are equal to the maximum weight in the network. Finally, $\tilde{C}_{i,Z}=1$ 
if each "outer" edge $w_{jk}=max(w)$, irrespective of the weights $w_{ij}$ of edges 
emanating from $i$.

\item Global $max(w)$ is used in normalization. This is true for all versions except 
$\tilde{C}_{i,B}$, where only the local strength $s_i$ matters.
This particular choice means that within the same network, two vertices whose 
neighbourhood topology and relative weight configuration are similar can have the 
same values of $\tilde{C}_{B}$ even if all the weights in the neighborhood of one 
vertex are small and those in the neighborhood of the other are large. 

\item Weights of all edges of triangles in which $i$ participates are taken into 
account. This is true for $\tilde{C}_{i,O}$ and $\tilde{C}_{i,H}$. However, $\tilde{C}_{i,B}$ 
takes into account only the weights of the edges connected to $i$. 
When $C_i=1$ and all $w_{jk, j\neq k}$ are equal, $\tilde{C}_{i,Z}=w_{jk}$, i.e. it 
is insensitive to the weights $w_{ij}$.

\item Invariance to permutation of weights within a single triangle. 
This feature is present only in $\tilde{C}_{i,O}$, showing that it 
deals with the triangles as an entity.

\item Weights of edges not participating in triangles are taken into account. 
This is the case for all definitions except $\tilde{C}_{i,O}$, where such edges
only enter through the vertex degree $k$.
\end{enumerate}

\begin{figure}[t]
\includegraphics[width=0.65\linewidth]{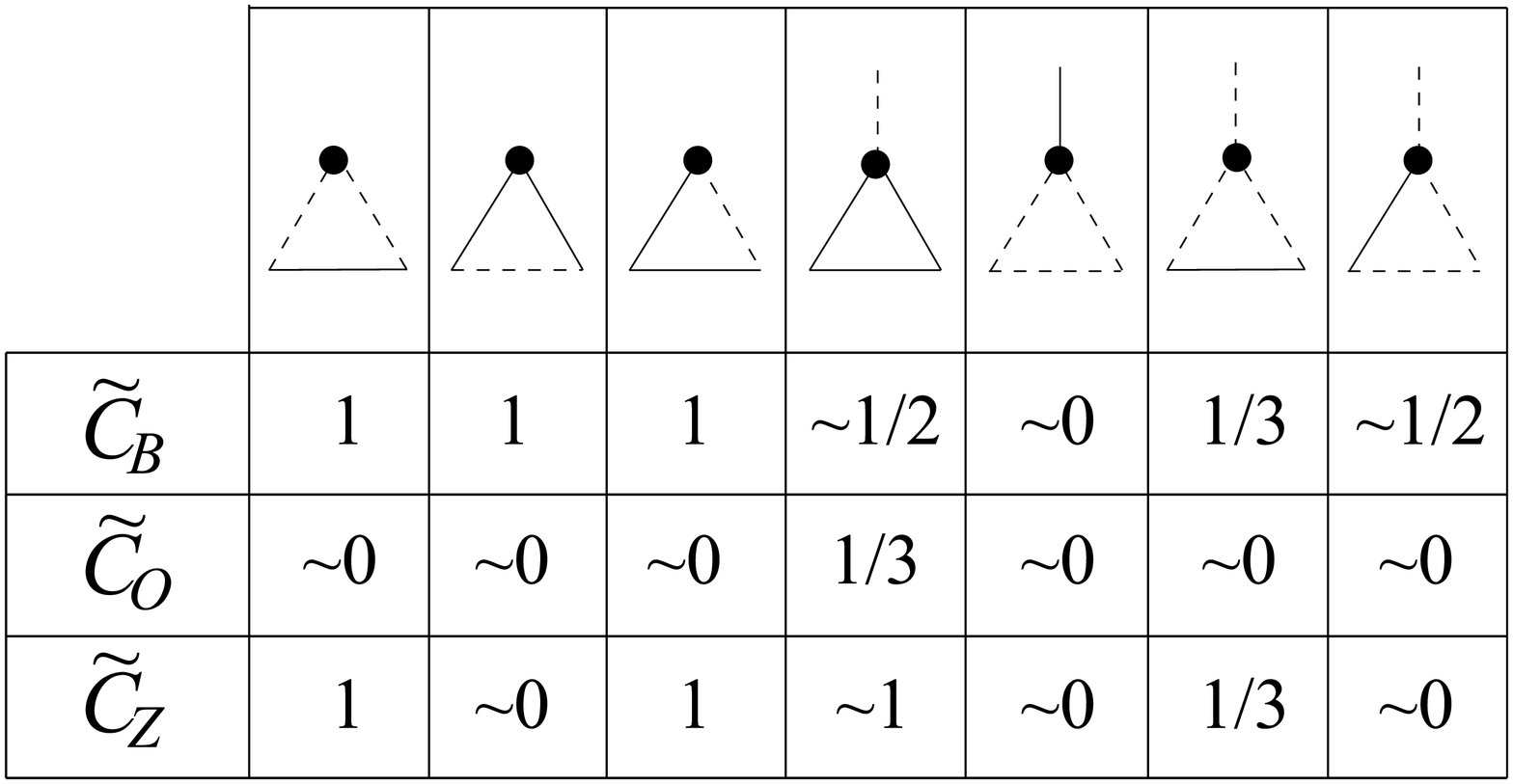}
\caption{Values of the weighted clustering coefficients for different weight configurations when vertex $i$ (solid circle) participates in a single triangle. Solid lines (---) depict edges of weight $w=max(w)=1$, whereas dashed lines ({\tiny - - -}) depict edges with vanishingly small weights $w=\epsilon \ll 1$. Note that in many cases different weight configurations yield the same coefficient values.}
\label{epsilon_plot}
\end{figure}

These differences are depicted also in Figure \ref{epsilon_plot}, which displays
the value of the clustering coefficient for a vertex participating in one triangle 
with varying weight configurations, including vanishingly small weights. 
In Fig.\ref{epsilon_plot} and in the following, 
analysis of $\tilde{C}_{i,H}$ is omitted as it is closely related to 
$\tilde{C}_{i,Z}$ but is normalized in a way which can be viewed as incorrect. 
Next, we compare the behavior of the different coefficients in two different empirical 
networks.

\begin{figure}[t!]
\begin{center}
\includegraphics[width=0.68\linewidth]{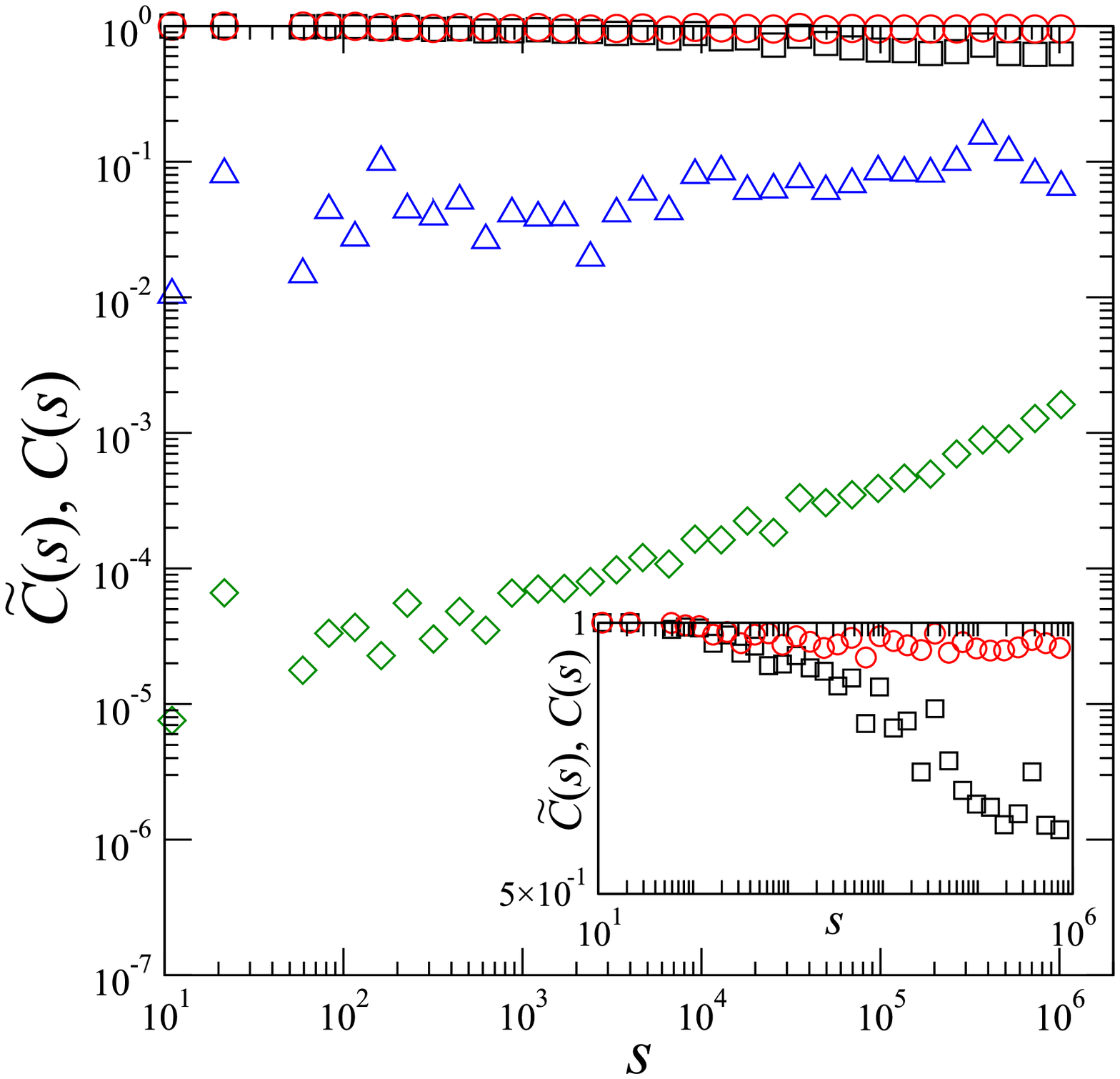}
\end{center}
\caption{Clustering coefficients computed for the international trade network 
(ITN) as function of vertex strength $s$: Unweighted $C(\square)$ and weighted   
$\tilde{C}_{B}$ $(\circ)$, $\tilde{C}_{O}$ $(\diamond)$, and $\tilde{C}_{Z}$ $(\bigtriangleup)$.
Inset: closer view on $C$ and $\tilde{C}_{B}$ with a linear vertical axis. } 
\label{fig:trade}
\end{figure}

\emph{-International Trade Network (ITN)}:
The ITN is constructed from trade records 
between the world's countries during the year 2000, such
that vertices denote countries, edges trade relationships, and edge weights trade volumes.
The source data \cite{itnsource} includes the dollar volumes of exports and imports
between countries but, due to different reporting procedures, there are usually small differences
between exports $exp_{ij}$ from $i$ to $j$ and imports $imp_{ji}$ to $j$ from $i$. 
We have chosen the edge weights $w_{ij}$ as a measure of the total trade volume such 
that $w_{ij}=\frac{1}{2}\left(exp_{ij}+exp_{ji}+imp_{ij}+imp_{ji}\right)$, averaging 
over the aforementioned discrepancies. The network constructed in this manner has 
$N=187$ countries connected with $E=10252$ edges, i.e., it has a relatively high 
edge density of 52 $\%$. 
High-trade-volume countries typically engage in high-volume trade with each other and, thus, in the network the high-weight edges are clustered, forming a ``rich-club''.

Figure \ref{fig:trade} depicts the different weighted clustering coefficients 
as function of 
vertex strength $s$. 
The unweighted clustering coefficient $C$ 
is also displayed for reference. Due to the large number of edges, 
$C$ remains high for all $s$. For low $s$, $\tilde{C}_{B}$ 
follows $C$ very closely, whereas 
for high values of $s$ $\tilde{C}_{B}$ gets values 
higher than $C$, which can be attributed to high-trade-volume countries engaging ´
in mutual high-volume trade. This effect is far more pronounced in  
$\tilde{C}_{O}$, which displays a power-law like increasing trend 
$\tilde{C}_{O}(s) \propto s^{\beta}$ with $\beta \approx 0.4$, spanning several 
decades. This effect is almost purely due to the behavior of the 
average triangle intensity $\bar{I}$ (see Eq.~(\ref{avg_int})), 
as the unweighted $C$ changes only a little. $\tilde{C}_{Z}$ is
seen to remain rather insensitive to the weights. Note that the overall 
level of $\tilde{C}_{O}$ and $\tilde{C}_{Z}$ is much lower than that of 
$C$ and $\tilde{C}_{B}$ due to weight normalization by the global $max(w)$ 
and to a broad distribution of weights.

\begin{figure}[t!]
\begin{center}
\includegraphics[width=0.75\linewidth]{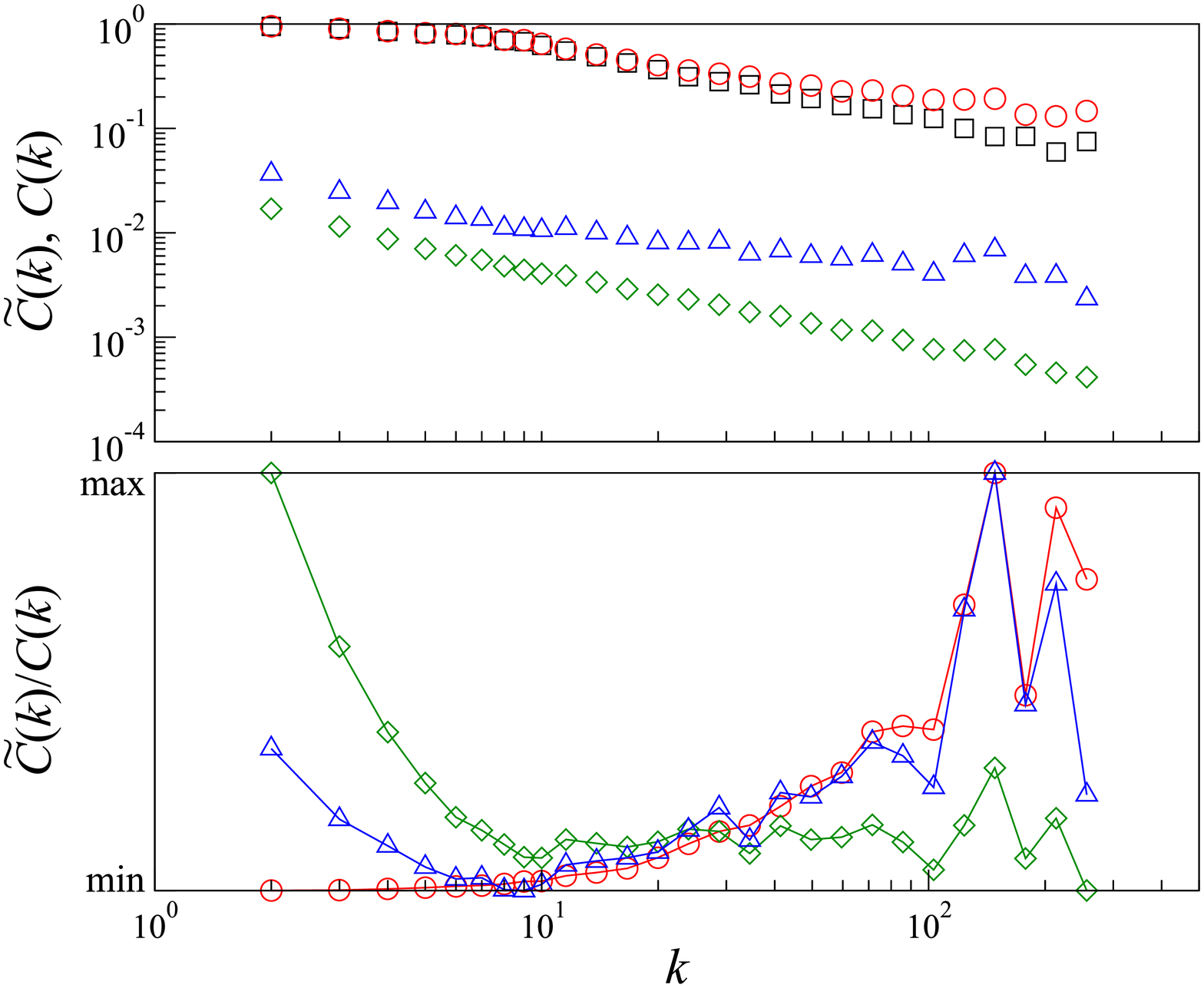}
\end{center}
\caption{Clustering coefficients computed for the scientific collaboration network (SCN) as 
function of vertex degree $k$:
unweighted $C$ $(\square)$, $\tilde{C}_{B}$ $(\circ)$, $\tilde{C}_{O}$ $(\diamond)$, and $\tilde{C}_{Z}$ $(\bigtriangleup)$. 
The lower panel displays the ratio of the weighted coefficients to the unweighted $C$; each curve
has been linearly scaled between its minimum and maximum values to facilitate comparison.}
\label{fig:SCN}
\end{figure}

\emph{-Scientific Collaboration Network (SCN)}: 
The SCN is constructed from scientists~\cite{NewmanColl} who have jointly 
authored manuscripts submitted to the condensed matter physics e-print 
archive (http://www.arxiv.org) from 1995 to 2005. In this network, vertices 
correspond to scientists and edges to co-authorships of papers. The edge weights have 
been defined such that 
$w_{ij} = \sum_p \left(\delta_{i,p} \delta_{j,p}\right)/\left(n_p-1\right)$, 
where the index $p$ runs over all the papers, $\delta_{i,p}=1$ if scientist 
$i$ is an author of paper $p$ and $0$ otherwise, and $n_p$ is the number of 
authors of paper $p$~\cite{NewmanColl}. This network has $N=40422$ nodes 
and an average degree of $\left<k\right>\approx 8.7$.

Figure~\ref{fig:SCN} displays different clustering coefficients as function 
of degree (upper panel) as well as the ratio of the weighted clustering coefficients 
to the unweighted coefficient (lower panel). Similarly to \cite{cbarrat} 
$\tilde{C}_{B}(k)$ remains rather close to $C(k)$ for $k<10$ but for $k>10$ 
their ratio is somewhat increased, indicating that the weights of edges that do
not participate in triangles are relatively low and/or the weights of 
edges participating in several triangles are relatively high.
In contrast, the shape of $\tilde{C}_{O}(k)$ differs from $C(k)$ for $k<10$. 
According to Eq.~(\ref{avg_int}), the ratio $\tilde{C}_{O}(k)/C(k)$ in 
the lower panel reflects the average intensity $\bar{I}(k)$ of triangles 
around vertices of degree $k$. The ratio is the largest for low-degree 
vertices, becoming approximately constant at $k\sim10$. A possible reason for 
this is that young scientists (e.g. graduate students) tend to participate 
in repeated collaborations involving a relatively small number of authors, 
giving rise to high-intensity triangles. $\tilde{C}_{Z}(k)$ appears to 
capture the low-$k$ behavior of $\tilde{C}_{O}$ as well as the high-$k$-behavior 
of $\tilde{C}_{B}$.

It is clear from the above considerations that there is no 
ultimate formulation for a weighted clustering coefficient.
Instead, we have seen that the different definitions capture different aspects 
of the problem at hand. For unweighted networks, it is straightforward 
to measure how many edges out of possible ones exist in the neighborhood of a 
vertex; yet the questions of how to measure the amount of weight located in this 
neighbourhood and what to compare this with, are far from obvious. 
In a sense $\tilde{C}_B$ and $\tilde{C}_O$ can be seen as limiting cases: 
$\tilde{C}_B$ compares the weights associated with triangles to the average weight 
of edges connected to the focal vertex, while $\tilde{C}_{O}$ 
disregards the strength of the focal node and measures triangle weights only 
in relation to the maximum edge weight in the network. $\tilde{C}_{Z}$ 
can be viewed as an interpolation between these two, 
albeit being a somewhat uncontrollable one as is evident from the examples 
in Fig.~\ref{epsilon_plot}. Given these observations, our conclusion is that there is 
no single general-purpose measure for characterizing clustering in weighted 
complex networks. Instead, it might be more beneficial to approach the problem 
from two angles. While the topological aspect can be described by the unweighted 
clustering coefficient $C$, the importance of the triangles can be quantified 
using the average triangle intensities of Eq.~(\ref{avg_int}). 

Support from the Academy of Finland (Center of Excellence program 2006-2011), 
OTKA K-60456 and COST P10 is acknowledged. We thank S.S.~Manna for discussions 
related to the ITN.


\begin{thebibliography}{99}
\bibitem{footnote:undirected} In this paper, we deal only with undirected networks.
\end{thebibliography}
\end{document}